\documentclass[twoside]{article} 
\usepackage{epsf,amsmath,amsfonts} 
\begin{document} 

\title{\vspace{-1in}\parbox{\linewidth}{\footnotesize\noindent
{\sc  Electronic Journal of Differential Equations},
Vol. {\bf 2001}(2001), No. 62, pp. 1-17. \newline
ISSN: 1072-6691. URL: http://ejde.math.swt.edu or http://ejde.math.unt.edu
\newline ftp  ejde.math.swt.edu  (login: ftp)}
\vspace{\bigskipamount} \\
Monotone Solutions of a Nonautonomous Differential Equation
for a Sedimenting Sphere
\thanks{  
\hfil\break\indent
\copyright 2001 Southwest Texas State University. \hfil\break\indent
Submitted XX. Published XX.} }

\date{}

\author{Andrew Belmonte, Jon Jacobsen, \& Anandhan Jayaraman}

\maketitle 
 
\begin{abstract} 
We study a class of integrodifferential  equations
and related ordinary differential equations for the  initial value 
problem of a rigid sphere falling through an infinite  fluid medium.  
We prove that   for creeping \textit{Newtonian} flow,
the motion of the sphere is monotone in its  approach to the
steady state solution given by the Stokes drag. We discuss this property in
terms of a general nonautonomous second order differential equation,
focusing on a decaying nonautonomous term motivated by 
the sedimenting sphere problem. 
\end{abstract} 

\newtheorem{definition}{Definition}[section]
\newtheorem{theorem}[definition]{Theorem}
\newtheorem{corollary}[definition]{Corollary}

\renewcommand{\theequation}{\arabic{section}.\arabic{equation}}
\catcode`@=11
\@addtoreset{equation}{section}
\catcode`@=12

\section{Introduction}
 
A rigid sphere falling through a viscous medium is a classic problem
in fluid dynamics, which was first solved in the steady state for 
the limit of vanishingly small Reynolds number in an infinite  domain
by G. G. Stokes in 1851 \cite{stokes:51}.   The time-dependent
approach to the steady state allows the partial differential equation
for the  sphere and fluid to be reduced to an integrodifferential
equation  for the motion of the sphere in  an infinite medium. The
physical effects included in this equation are the  buoyancy which
drives the motion, the inertia of the sphere, the  viscous drag, an
added mass term, and a memory or history  integral \cite{clift}.

In the case of a {\it Newtonian fluid}, the main effect of the memory 
integral on the dynamics is to modify the approach to steady state 
from   exponential to algebraic.  The integral also makes the 
equation effectively second order, though it is generally accepted 
that no oscillations occur as the sphere reaches its steady state 
value   \cite{clift,walt92}. Physically it is clear that no
oscillations can occur due to the absence  of a restoring force
against gravity, and a sphere   released from  rest in a Newtonian
fluid at low Reynolds number is observed to reach its terminal 
velocity monotonically. However, it is not directly evident
mathematically that oscillating solutions are precluded,
particularly as the governing integrodifferential equation can be
transformed to a  nonautonomous second order ordinary differential
equation which has  the form of a harmonic oscillator
\cite{clift,villat:44}.

In a {\it non-Newtonian fluid}, such as a polymer solution,  a falling
sphere is often observed to  undergo transient  oscillations before
reaching its terminal velocity  \cite{arigo97,walt92}. These 
oscillations occur due to the elasticity of the fluid, which provides
a restoring force \cite{bird87}. The steady state value is of primary
concern  in many applications, and much work focuses only on this
aspect of  the problem. The oscillations which occur during the
approach to  steady state have been reproduced in a linear
viscoelastic model by King and Waters \cite{king:ums72}. 
 
More recently, \textit{nontransient} oscillations of falling spheres
(and  rising bubbles)
have been observed in specific aqueous solutions
of surfactants (wormlike micellar solutions) 
\cite{belmonte:soc00,jayaraman:osf00}. These observations were
initially made for a bubble in the wormlike micellar fluid CTAB/NaSal
\cite{hu98,rehage82}, which showed oscillations in its position and
shape.  The shape
oscillations included an apparent cusp which momentarily appears at
the trailing end of the bubble. Such a cusplike tail is a well known
property of rising bubbles in non-Newtonian fluids
\cite{hass79,liu95}, which we initially believed to play an important
role in the micellar oscillations. 

%%%%%%%%%%%%%%%%%%%%%%%%%%%%%%%%%%%%%%%%%%%%%%%%%%%%%%%%%%%%%%%%%%%%%%% 
 
\epsfxsize=5.5cm  
\begin{figure}                                                                                                                    
\begin{center} 
\[  
\epsfbox{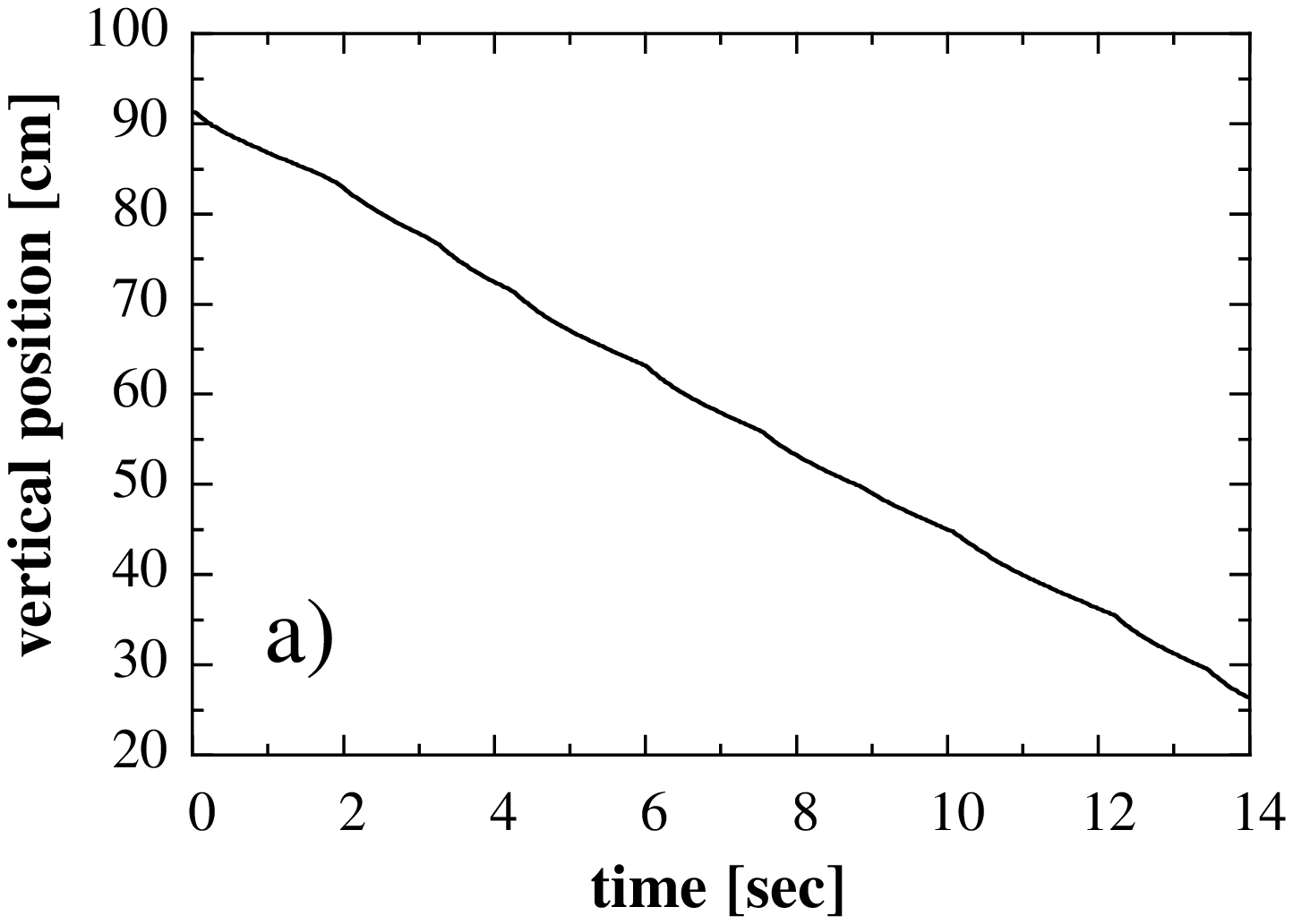} 
%\epsfbox{1_8spher.eps} 
\epsfxsize=0.6cm  
\epsfbox{blank.eps} 
\epsfxsize=5.5cm  
\epsfbox{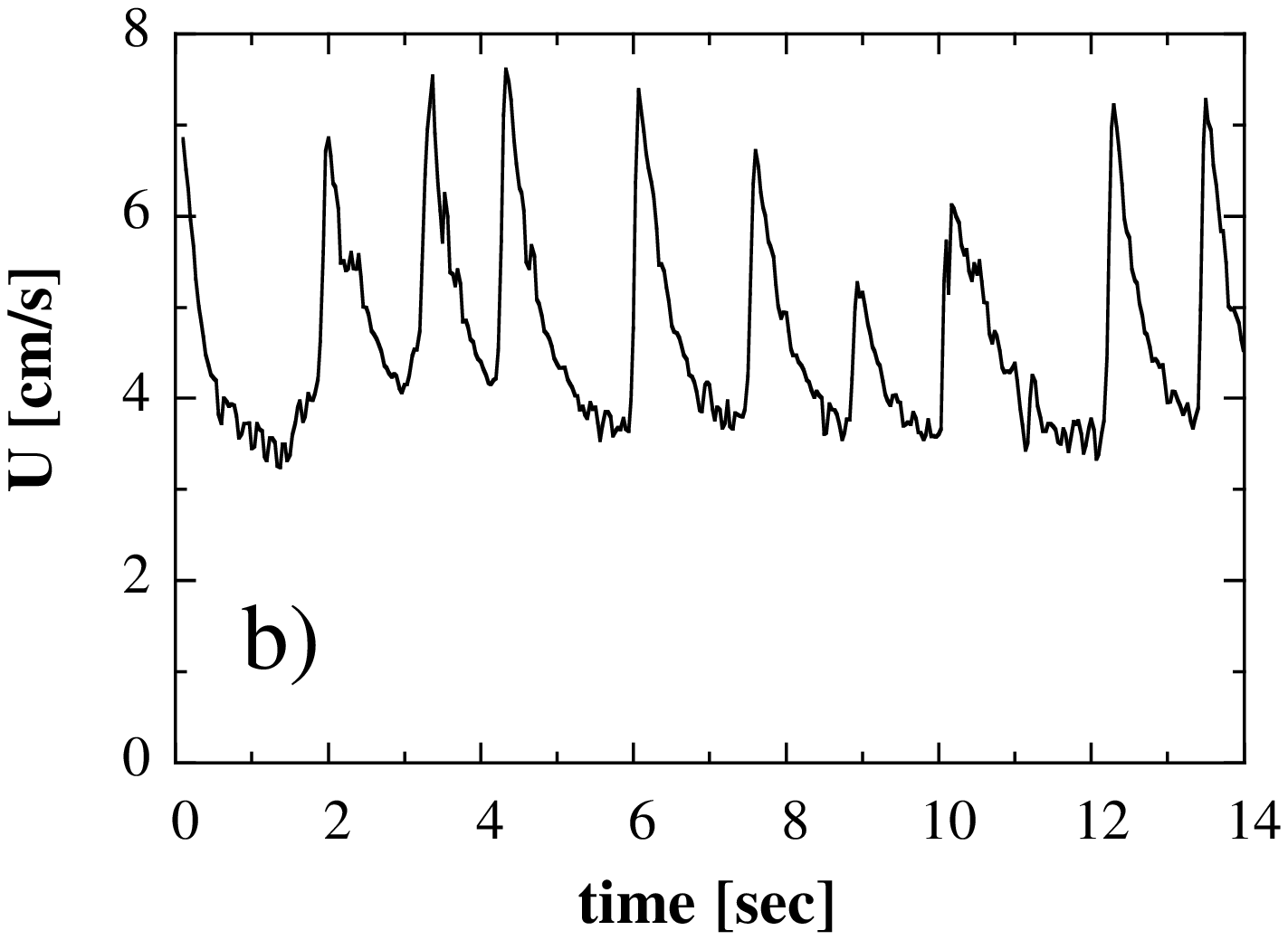} 
 \]                                       
\end{center} 
                                                                                                       
\caption{Motion of a 1/8'' diameter teflon sphere falling through an  
aqueous solution of 6 mM CTAB/NaSal: a) position vs time; b)  
calculated velocity vs time.} 
\label{f-expt}                                                                                                                   
\end{figure}                                                                     
 
%%%%%%%%%%%%%%%%%%%%%%%%%%%%%%%%%%%%%%%%%%%%%%%%%%%%%%%%%%%%%%%%%%%%%%%  

Subsequent observations of solid spheres which also oscillate while  
falling through the same solutions made it   clear that the cusp is
not involved in the phenomenon, and that   another explanation must
be sought. Unlike a sedimenting sphere in a conventional
non-Newtonian fluid, these oscillations do not appear to be transient
\cite{jayaraman:osf00}.  An example is given
in Figure \ref{f-expt}, which shows the motion of a 1/8''   teflon
sphere falling through a tube ($L = 98$ cm, $R = 3.2$ cm)   filled
with a 6 mM 1:1 solution of CTAB/NaSal \cite{jayaraman:osf00}.

Our attempts to model this phenomenon  brought to our attention the
unusual aspects of the  integrodifferential equation for a falling
sphere. We prove here that the equation for sedimenting sphere in a
Newtonian fluid in the limit of zero Reynolds number (creeping flow)
does not  admit   oscillating solutions, despite some appearances
that it  does.  This result is due to the special  properties of the
error   function when multiplied by oscillating  functions.  It is
ultimately   related to the stability of  nonautonomous ordinary
differential equations with monotone secular terms,  which is
appropriately  viewed as an initial value problem, and not in terms
of linear stability analysis around the terminal velocity.

\section{The Motion of a Sedimenting Sphere }
%in a Newtonian Fluid} 
 
We begin by reviewing some classical results for
the equation of motion governing a falling sphere in
a viscous Newtonian fluid of infinite extent (no boundaries).

\subsection{Equation of Motion of the Sphere}
An incompressible fluid in the absence of body forces is described 
by the equations
\begin{eqnarray}
\label{ce2}
\rho \left( \frac{\partial \vec u}{\partial t} + (\vec u \cdot 
\nabla) \vec u \right) & = & -\nabla p +  \, \text{div} \, \sigma, \\
                \text{div} \, \vec u & = & 0,
\label{ce3}
\end{eqnarray}
where $\rho(\vec x,t)$ is
the density of the fluid, $p$ is the  pressure,
$\vec u(\vec x,t)$ is the velocity field for the fluid, and
$\sigma(\vec x,t)$ is the extra stress tensor, which measures 
force per unit area (other than pressure) in the
present configuration of the fluid.  
A Newtonian fluid is a fluid for which the stress tensor $\sigma$ is
linearly related to the rate of strain tensor $D$ through the relation
\begin{equation}
   \sigma = 2 \mu D,  
\label{st}
\end{equation}
where  $\mu$ is the viscosity of the fluid 
and $D = (\nabla \vec u + (\nabla \vec u)^T)/2$ is the symmetric part of
the velocity gradient $\nabla \vec u$.   
From (\ref{ce2}),(\ref{ce3}), and (\ref{st}) one
obtains the Navier-Stokes equation:
\begin{equation}
\rho \left(\frac{\partial \vec u}{\partial t} + (\vec u \cdot 
\nabla) \vec u \right)  =  -\nabla p + \mu \Delta \vec u.
\label{NS}
\end{equation}

Non-Newtonian fluids are fluids for which the assumption (\ref{st}) is
invalid.  For instance, polymeric and viscoelastic fluids often fail to
conform to the instantaneous relation between stress and velocity gradients
implicit in (\ref{st}). In general $\sigma=\sigma(t,D)$ will depend nonlinearly
on $D$ and on the past history of stress in the fluid.

By choosing a time scale and an appropriate length scale, (\ref{NS}) can 
be written in a nondimensional form
\begin{eqnarray}
 \frac{\partial \tilde u}{\partial \tau} +\mbox{Re}\; (\tilde u \cdot 
\nabla) \tilde u  & = & -\nabla \tilde p +  \Delta \tilde u \label{NS2} \\
                \text{div} \, \tilde u & = & 0, \label{incomp2}
\end{eqnarray}
where, $\tilde u$ and $\tilde p$ are the nondimensional velocity and pressure. 
The dimensionless constant $\mbox{Re}$ is called the Reynolds number and it 
measures the relative importance of inertial effects to that of viscous effects.
When the inertial effects are negligible ($\mbox{Re}=0$), equation
(\ref{NS}) is called
the Stokes equation. In this paper we restrict our analysis to this situation.

Stokes solved the steady version of  (\ref{NS2})-(\ref{incomp2}) for the case 
of sphere falling though the fluid for vanishing Reynolds number. The Stokes 
solution gives the steady state drag on the sphere of radius $R$ falling 
through a fluid with a steady speed $U$ to be $F=6 \pi \mu R U$.
However, in order to solve the transient problem of falling sphere, we first 
solve the problem of sphere oscillating with a frequency $\omega$ and compute 
the drag experienced by the sphere as a function of $\omega$.  The drag experienced 
by a sphere falling at a arbitrary speed $U(t)$ can then be computed as a Fourier 
integral of this drag.

Consider a sphere of radius $R$ and density $\rho_s$ in a Newtonian
fluid of density $\rho$ and viscosity $\mu$.  The exterior Stokes
flow driven by small oscillations of the sphere at a frequency
$\omega$ can be solved exactly \cite{basset,lamb}, which leads to a
hydrodynamic force dependent on both $U$ and $dU/dt$ : 
\begin{eqnarray} 
F=6 \pi \mu R\left(1+\frac{R}{\delta}\right) U + 3 \pi R^2 \rho \delta 
\left(1+\frac{2R}{9\delta}\right) \frac{dU}{dt},\label{eq:Fw-drag} 
\end{eqnarray} 
where $\delta=\sqrt{2\nu/\omega}$ is a diffusive lengthscale common  
to Stokes problems, and $\nu = \mu/\rho$ is the kinematic viscosity.
  Using this, the
general time-dependent problem of the motion of a falling sphere can
be reduced from a partial to an ordinary differential equation for
the speed $U(t)$ of the sphere, an exact equation which takes into
account the motion of the surrounding fluid \cite{clift,villat:44}.  
For a sphere moving with an arbitrary speed $U(t)$, the 
hydrodynamic drag it experiences can be calculated by representing 
$U(t)$ as a Fourier integral: 
$$ 
U(t)= \int^\infty_{-\infty} U_\omega e^{-i \omega t}\; d\omega.  
$$ 
The drag for each Fourier component is then given by  
(\ref{eq:Fw-drag}).  The total hydrodynamic drag on the sphere is  
obtained by integrating over all Fourier components, leading to 
\begin{eqnarray} 
F_{drag} = 6\pi\mu R U(t) + \frac{1}{2} \rho \mathcal{V} \frac{dU}{dt}  
+ 
 6\pi\rho R^2 \sqrt{\frac{\nu}{\pi}} \int^t_{-\infty}  
\frac{U'(s)}{\sqrt{t-s}}\;ds \label{eq:F-net} 
\end{eqnarray} 
where
the first term represents the  
steady state drag on a sphere falling with a velocity $U$,  
the second term represents the added mass term (the force  
required to accelerate the surrounding fluid), the  
third term is the Basset memory term, and
 $\mathcal{V}=4 \pi R^3/3$ is the volume of the sphere. 
If the sphere starts from rest, then the lower limit of the integral  
in (\ref{eq:F-net}) starts from $\tau=0$ instead of  
$\tau=-\infty$.  The expression for the unsteady drag force can then  
be substituted into the balance of force equation for the sphere: 
$$ 
\rho_s \mathcal{V} U'(t) = F_{buoy} - F_{drag}. 
$$ 
Thus the equation of motion for the sphere is 
\begin{eqnarray} 
(\rho_s+\frac{\rho}{2}) \mathcal{V} U'(t) \; 
&&+ \; 6\pi \mu R U(t) \,+ \;6\pi R^{2}\sqrt{\frac{\rho\mu}{\pi}} \int^t_0 
\frac{U'(s)}{\sqrt{t-s}}\, ds  \nonumber\\ 
&&  
= \, (\rho_s-\rho) \mathcal{V} g, 
\label{eq-forces} 
\end{eqnarray} 
which can be rewritten in the simpler form 
\begin{equation} 
U'(t) + B U(t) + Q \int^t_0 \frac{U'(s)}{\sqrt{t-s}}\, ds  
 =M, \label{eq:uprime} 
\end{equation} 
where 
\begin{eqnarray} 
B&=&\frac{9\mu}{R^2\left(2\rho_s + \rho\right)}, \label{eq:B} \\ 
Q&=&\frac{9\rho}{R\left(2\rho_s+\rho\right)}  
\sqrt{\frac{\mu}{\rho\pi}}, \label{eq:Q} \\ 
M&=&\frac{2 g \Delta \rho}{2\rho_s + \rho}, \label{eq:M} 
\end{eqnarray} 
and $\Delta\rho = \rho_s - \rho$ is the density difference which 
drives  the motion.  In this approach the motion of the sphere is 
described by an integrodifferential equation whose
integral term has the same singularity as Abel's equation 
\cite{keener:pam00,estrada:sie00}. 
Note that this equation is only valid 
in the limit of zero Reynolds number \cite{ock68,michael97}. 
  
Physically one expects the solution $U(t)$ to approach  
a terminal velocity.  It is clear from 
(\ref{eq:uprime}) that the only steady state solution ($U'=0$)  
possible is  
\begin{equation} 
U_0 = \frac{M}{B} = \frac{2\Delta\rho gR^2}{9\mu},  
\label{e-stokes} 
\end{equation} 
which is the classical result of balancing the Stokes drag with the  
buoyancy. 
 
\subsection{Solving the Integrodifferential Equation} \label{solnIDE} 
 
We first rewrite the  integrodifferential equation (IDE) for the 
sphere in a  nondimensional form using $U_0$ as the velocity scale, 
and $1/B$, the viscous diffusion time, as  the time scale. The 
variables are 
$$ 
\tau = Bt \qquad \text{and} \qquad u(\tau) = U(\tau/B )/U_0. 
$$ 
With this rescaling (\ref{eq:uprime}) becomes
\begin{equation} 
u'(\tau) + u(\tau) + \sqrt{\frac{\kappa}{\pi}}\int^{\tau}_0  
\frac{u'(s)}{\sqrt{\tau-s}}\, ds =1, \label{eq:maineq} 
\end{equation} 
where the control parameter $\kappa$ is given by 
\begin{equation} 
\kappa = \frac{\pi Q^2}{B}=\frac{9\rho}{2\rho_s+\rho}. 
\label{eq:kappa} 
\end{equation}
Thus the motion of the sphere depends only on the relative densities 
of  the sphere and the fluid through the parameter $\kappa$. 
The  density of the sphere $\rho_s$ can range from zero to infinity,
which implies a parameter range  $0 < \kappa < 9$. When the density
of the sphere is  equal to the density of the fluid,  $\kappa=3$ and
$U_{0}=0$. Here we will only be concerned with falling spheres, for
which $0 < \kappa < 3$.

Although we will solve the IDE (\ref{eq:maineq}) directly, it is of  
interest to connect the problem to ordinary differential equations  
(ODE's) and discuss some important consequences therein, especially  
with regard to the stability of the terminal velocity solution.  
Following Villat \cite{villat:44}, we can  
rewrite (\ref{eq:maineq}) as an ODE using Abel's Theorem (see Appendix
\ref{app}):
\begin{eqnarray} 
u'' + (2-\kappa) u' + u = 1- \sqrt{\frac{\kappa}{\pi  
\tau}},  
\label{eq:non-dim1} \\ 
u(0) = 0, \quad u'(0) = 1.   
\label{eq:non-dim2} 
\end{eqnarray} 
More general initial conditions $U(0) \neq 0$ and $U'(0)=M -  
BU(0)$ lead to a slightly different ODE: 

\begin{eqnarray} 
u'' + (2-\kappa) u' + u & = & 1 + \sqrt{\frac{\kappa}{\pi  
\tau}}(u(0) - 1), 
\label{eq:gen1} \\ 
u(0) & = & \xi, \quad u'(0)  =  1 - \xi. \label{eq:gen2}  
\end{eqnarray} 
Note that it follows from equation (\ref{eq:maineq}) that $u'(0)$ is
prescribed in terms of $u(0)$, and thus the second order ODE we have
obtained requires only one initial condition.

Since we are investigating the possibility of steady-state
oscillations of a sedimenting sphere, we are primarily concerned with
the asymptotic behavior of (\ref{eq:non-dim1}). Moreover, since the
nonautonomous term tends to zero as $t\rightarrow\infty$, one might
expect the  stability of (\ref{eq:non-dim1}) to mimic the homogeneous
problem. With this in mind, let $\alpha$ and $\beta$ denote the roots
of the characteristic equation 
\begin{equation}
m^2+(2-\kappa)m+1 =0. 
\label{ce}
\end{equation}
It is readily verified that (\ref{ce}) has complex roots
for  $0 < \kappa < 4$.  Moreover, the roots have positive real parts
for   $2 < \kappa < 4$.  
Since the relevant range of $\kappa$ for a falling
sphere is $0 < \kappa < 3$, one sees that oscillations are
not a priori precluded.
In terms of the actual densities of the fluid and  sphere
the condition for complex roots corresponds to  $\rho_s > (5/8) \rho$, 
which is true in the case of a heavy sphere falling through a lighter 
liquid ($\rho_s > \rho$).  If additionally
$\rho_s < (7/4) \rho$, then the  complex roots have positive real parts.
If we rewrite (\ref{eq:non-dim1}) in the asymptotic limit ($t \rightarrow \infty$)
as a  first  order linear system
\begin{eqnarray} 
x' &=& y \\ 
y' &=& (1-x) + (\kappa-2) y, 
\end{eqnarray}
where $x=u$, and $y = u'$, then 
$(x,y)=(1,0)$ is the unique equilibrium point,
which corresponds to the terminal velocity.  The  
eigenvalues of this  system are precisely $\alpha$ and   $\beta$, whence
the equilibrium  point becomes unstable.

Nonetheless, as we will show, even in this range ($2 < \kappa < 3$), the solution 
to the full equation (\ref{eq:non-dim1}) \textit{monotonically} approaches
the value 1, corresponding to the monotonic 
approach to the steady Stokes value (\ref{e-stokes}) for the actual velocity $U$.    
Clearly the nonautonomous term continues to play a dominant role in the stability
of (\ref{eq:non-dim1}), despite its algebraic approach to zero.

To solve for $u=u(\tau)$, we return to the IDE (\ref{eq:maineq}) and apply the Laplace  
transform in the case $u(0) = 0$:  
\[ 
\mathcal{L}\{u\}(s) = \frac{1}{s(s + \sqrt{\kappa} \sqrt{s} + 1)} \quad   
\text{or} \quad 
\mathcal{L}\{u'\}(s) = s \mathcal{L}\{u\} = \frac{1}{s + \sqrt{\kappa} \sqrt{s} + 1} . 
\] 
Since 
 $\alpha \beta = 1$ and $ \sqrt{\alpha}+ \sqrt{\beta} = \sqrt{\kappa} $, we may 
express this last equation in the form 
\[   
\mathcal{L}\{u'\}(s)  = \frac{\sqrt{\kappa}}{\alpha - \beta}  
\left[\frac{\sqrt{\alpha}}{\sqrt{s}(\sqrt{s} + \sqrt{\alpha})} 
- \frac{\sqrt{\beta}}{\sqrt{s}(\sqrt{s} + \sqrt{\beta})}\right]. 
\] 
Moreover, the identity  
\[   
 \mathcal{L}\{e^{\alpha t} \text{Erfc} \sqrt{\alpha t} \} = \frac{1}{\sqrt{s}(\sqrt{s} +  
\sqrt{\alpha})},   
\] 
implies 
\begin{equation} 
   u'(\tau) = \frac{\sqrt{\kappa}}{\alpha - \beta} 
                \left[\sqrt{\alpha} e^{\alpha \tau} \text{Erfc} \sqrt{\alpha \tau} -  
                \sqrt{\beta} e^{\beta \tau} \text{Erfc} \sqrt{\beta \tau} \right].   
  \label{eq:uprime2} 
\end{equation} 
Finally, since  
\[  \sqrt{\alpha} \int e^{\alpha t} \text{Erfc} \sqrt{\alpha t} \, dt = 2 \sqrt{\frac{t}{  
\pi}} +  
\frac{1}{\sqrt{\alpha}} e^{\alpha t} \text{Erfc} \sqrt{\alpha t} + C,  
\] 
we find the solution  
\begin{eqnarray}  
u(\tau) = 1 + \frac{\sqrt{\kappa}}{\alpha -\beta} 
\left[ \frac{e^{\alpha \tau} \text{Erfc} \sqrt{\alpha \tau}}{\sqrt{\alpha}}  -   
\frac{e^{\beta \tau}\text{Erfc} \sqrt{\beta \tau}}{\sqrt{\beta}}  
\right]. 
\label{eq:soln} 
\end{eqnarray} 
 
This solution to the IDE (\ref{eq:maineq}) is 
also the solution to the ODE (\ref{eq:non-dim1})-(\ref{eq:non-dim2}). 
Applying transform methods to the more general set of equations defined by 
(\ref{eq:gen1})-(\ref{eq:gen2}) one finds the solution
\begin{equation}    
u(\tau) = (1-\epsilon) u_0(\tau) 
+ \epsilon\, \qquad \epsilon \equiv u(0),\, u'(0)=1-\epsilon,
\label{220}
\end{equation}
where $u_0(\tau)$ is the solution defined by (\ref{eq:soln}).  Note that
the solution for arbitrary initial velocity $\epsilon$ is a simple
rescaling of the solution for the sphere  initially at rest.  It is not
obvious from the form of $u$ in (\ref{eq:soln}) that the values approach 1 monotonically.
Let us first investigate the asymptotic behavior of this solution.

\subsection{Asymptotic Approach to the Steady Stokes Solution} 
\label{asympt} 
 
Finding the asymptotic behavior of the solution $u_0$ is straightforward.
We employ the asymptotic expansion of the error function
\cite{gautschi:eff92} 
\begin{eqnarray} 
e^{z^2} \text{Erfc} \, z  \sim \frac{1}{\sqrt{\pi}z}\left[1 + 
  \sum_{m=1}^\infty (-1)^m \frac{1\cdot3 \cdots  
(2m-1)}{(2z^2)^m}\right] 
\end{eqnarray} 
as $z\rightarrow\infty$, provided $|\arg (z)| < 3\pi/4$. 
Since $0 < \arg(\sqrt{\alpha t}) < \pi/2$ for $0 < \kappa < 3$, we  
see that asymptotically  
$$ 
\text{Erfc} \sqrt{\alpha \tau} \sim \frac{e^{-\alpha \tau}}{\sqrt{\alpha  
\tau}}. 
$$ 
 
Thus the product of the exponential term and the error function  
approaches zero in the limit $t\rightarrow\infty$. Using this  
expansion in (\ref{eq:soln}) we obtain 
\begin{eqnarray} 
 \lim_{\tau \rightarrow \infty} u_0(\tau) = 1 \quad \text{ or }  
 \quad \lim_{t \rightarrow \infty} U(t) = U_0. 
\end{eqnarray} 
As the solution for any initial condition is a 
rescaling of $u_0$, we see this limit applies for all values of $u(0)$. 
 
Although the asymptotics of this solution are clear, the transient
solution has some unusual properties. Numerical
simulation of the IDE (\ref{eq:maineq}), or even attempts to plot the
analytic solution (\ref{eq:soln}), eventually  blow up at
large $\tau$ when $\kappa$ is in the unstable range. 
Clearly the cancellation between the exponentially growing and 
decreasing terms are quite sensitive to numerical errors.  This is an  
indication that the product $e^{\alpha \tau}\text{Erfc} \sqrt{\alpha
\tau}$ should be considered as a special function with its own
properties.

\section{Monotonicity of the Transient Solution} 
\label{monosection}
 
We begin with the transient solution to the IDE or ODE considered
in   the previous section.  In addition to the insensitivity of the  
transient solution to the real part of the homogeneous roots, it is  
surprising that the nonzero complex part  does not lead to
{\it any} oscillations in the velocity of the sphere, although there
has sometimes been confusion on this point regarding transient
oscillations   \cite{villat:44}.  Experimentally the sphere in a
Newtonian fluid has never   been observed to oscillate, in
contradistinction to most non-Newtonian   (particularly elastic)
fluids \cite{arigo97,walt92}. We will show   that the solution $u$
defined by (\ref{eq:soln}) is monotone as a   function of $\tau$ for all
$\kappa \in (0,4)$.  Although this may be well   known, we have not yet
found a reference to any proof other than the case  
$\alpha,\beta \in \mathbb{R}$ \cite{wil78}, which 
corresponds to $\kappa > 4$.

Let us define the function $\text{Vi}:\mathbb{C} \rightarrow \mathbb{C}$ by
\begin{equation} 
\text{Vi}(z) \doteq e^{z}\text{Erfc} \sqrt{z}  
= \frac{2e^{z}}{\sqrt{\pi}} \int_{\sqrt{z}}^{\infty} e^{-s^{2}}\,ds. 
\label{defvi}
\end{equation}
We shall refer to $\text{Vi}$ as the Villat function, since this combination 
appeared in the explicit solution of the differential equation for the falling 
sphere problem by Villat \cite{villat:44}.  Closely related to $\text{Vi}(z)$ is
the ``plasma  dispersion  function'' $w$, defined by
\cite{gautschi:eff92}:  \[   w(z) = e^{-z^2} \text{Erfc} (-i z).  \]  In
fact, $\text{Vi}(\alpha t) = w(i \sqrt{\alpha t})$.   Using the Villat function
we may now prove the main theorem.

\begin{theorem}[Monotonicity] For each $\kappa \in (0,4)$, the function
\begin{equation}  
u(\tau) = 1 + \frac{\sqrt{\kappa}}{\alpha -\beta} 
\left[ \frac{e^{\alpha \tau} \text{Erfc} \sqrt{\alpha \tau}}{\sqrt{\alpha}}  -   
\frac{e^{\beta \tau}\text{Erfc} \sqrt{\beta \tau}}{\sqrt{\beta}}  
\right] 
\end{equation}
approaches the limit 1 monotonically.
\end{theorem} 

\paragraph{Proof.}
We have shown that $u(t) \rightarrow 1$, thus it remains to show it does so  
monotonically.  We will demonstrate this by proving $u'(t) > 0$ for  
all $t > 0$.  To this end, fix $t > 0$, $\kappa \in (0,4)$ and recall that $\alpha$ and  
$\beta$ denote the conjugate pair of roots of the polynomial $m^2 +  
(2-\kappa)m + 1$.  Recall from (\ref{eq:uprime2}) that 
\[   
u'(t) = \frac{\sqrt{\kappa}}{\alpha - \beta} 
        \left[ \sqrt{\alpha} \, \text{Vi}(\alpha t) - \sqrt{\beta} \, \text{Vi}(\beta t)  
\right].    
\] 
Since $\text{Erfc}(\overline{z}) = \overline{\text{Erfc} \, z}$, it follows that  
$\text{Vi}(\beta t) = \overline{\text{Vi} (\alpha t)}$ and 
\begin{equation} 
u'(t) = \sqrt{\kappa} \,  \frac{\sqrt{\alpha} \, \text{Vi}(\alpha t) -  
\overline{\sqrt{\alpha} \, \text{Vi}(\alpha t)}}{\alpha - \overline{\alpha}} 
= \sqrt{\kappa} \,  \frac{\Im \{\sqrt{\alpha} \, \text{Vi}(\alpha t) \}}{\Im \{\alpha \}}.       
\label{eq:uprimeform2} 
\end{equation} 
Since $\Im \{ \alpha \} > 0$ for each $\kappa \in (0, 4)$, it is evident from (\ref{eq:uprimeform2})  
that the sign of $u'(t)$ is determined by the imaginary part of  
the function $\sqrt{\alpha} \, \text{Vi}(\alpha t)$.   For the  plasma  
dispersion function $w$ introduced above, the real and imaginary parts are given by 
(see e.g., \cite[7.4.13-7.4.14]{gautschi:eff92}) 
\[  
\Re(w(x + i y)) = \frac{1}{\pi} \int_{-\infty}^{\infty} \frac{y  
e^{-s^2}}{(x-s)^2 + y^2} \, ds \quad (x \in \mathbb{R}, y > 0) 
\] 
and  
\[  
\Im(w(x + i y)) = \frac{1}{\pi} \int_{-\infty}^{\infty} \frac{(x-s)  
e^{-s^2}}{(x-s)^2 + y^2} \, ds \quad (x \in \mathbb{R}, y > 0). 
\] 
 It is readily verified that  
$| \alpha | = 1$, thus in polar form we have $\alpha = e^{i \theta}$ for  
some fixed $\theta \in (0,\pi)$, in which case $\text{Vi}(\alpha  
t) = w(i \sqrt{\alpha t}) = w(x + i y)$, where 
$$ 
x = -\sqrt{t} \sin \left( \frac{\theta}{2} \right) 
\qquad \text{and} \qquad y = \sqrt{t}  
\cos \left( \frac{\theta}{2}\right) . 
$$   
Using this information we compute 
\begin{eqnarray} 
 \Im \{\sqrt{\alpha} \, \text{Vi}(\alpha t)\} & = & 
\cos \left( \frac{\theta}{2}\right) \, \Im  
\{ w(x + i y) \} 
                                        + 
\sin \left( \frac{\theta}{2} \right) \,  
\Re \{ w(x + i y)\}  
\nonumber  \\ 
                & = & \frac{1}{\pi} \cos \left( \frac{\theta}{2} \right)  
\int_{-\infty}^{\infty} \frac{(x-s) e^{-s^2}}{(x-s)^2 + 
y^2} \, ds \nonumber \\
& & \qquad 
 + \frac{1}{\pi} \sin \left( \frac{\theta}{2} \right) 
\int_{-\infty}^{\infty}  
\frac{y e^{-s^2}}{(x-s)^2 + y^2} \, ds  
 \\ 
         & = & \frac{1}{\pi} \cos \left(  \frac{\theta}{2} \right) 
\int_{-\infty}^{\infty} \frac{-s e^{-s^2}}{(x-s)^2 + y^2} \, ds. 
\label{eq:poseq} 
\end{eqnarray} 
In the last step  we have used the fact that 
$x \cos(\theta/2) + y  \sin(\theta/2) = 0$. 
Since $\sqrt{\alpha}$ lies in the first quadrant, the prefactor of the 
last integral above is positive and we may conclude 
that  $\Im \{\sqrt{\alpha} \, \text{Vi}(\alpha t)\} > 0$ 
provided 
\begin{equation} 
\int_{-\infty}^{\infty} \frac{s e^{-s^2}}{(x-s)^2 + y^2} \, ds  
=  \int_{-\infty}^{\infty} \frac{s e^{-s^2}}{(s + \sqrt{t}  
\sin \frac{\theta}{2})^2 + t \cos^2 \frac{\theta}{2} } \, ds < 0.   
\label{negeq} 
\end{equation} 
Let us denote the integrand as  
\[ F(s) = \frac{s e^{-s^2}}{P(s)} \quad \text{where} \quad 
 P(s) = \left(s + \sqrt{t} \sin \frac{\theta}{2} \right)^2 + t  
\cos^2 \frac{\theta}{2}.  \] 
Note that $P(s) > 0$ for $s \in  
\mathbb{R}$ (recall $t > 0 $ is fixed).  The proof is complete once the 
following two  observations are made: 

\begin{enumerate} 
\item[($a$)] \, $| F(-s) | > F(s)$ for $s > 0$;  
\item[($b$)] \, $\int_\mathbb{R}  F \, ds = \int_0^{\infty} F(s) \, ds - \int_0^{\infty} |  
F(-s) | \, ds$.
\end{enumerate} 
To see $(a)$, notice for $s > 0$ we have $0 < P(-s) < P(s)$, thus 
$$ 
|F(-s)| = \frac{s e^{-s^2}}{P(-s)} > \frac{s e^{-s^2}}{P(s)} =  
F(s),  
\qquad \text{for }  s > 0. 
$$ 
Observation $(b)$ follows from a standard change of variables 
\[ 
\int_\mathbb{R} F(s) \, ds = \int_{-\infty}^0 F(s) \, ds + \int_0^{\infty}  
F(s) \, ds = \int_0^{\infty} F(s) \, ds - \int_0^{\infty} |  
F(-s)| \, ds .   
\]  
The two observations above imply the inequality (\ref{negeq}) holds, 
in which case by (\ref{eq:uprimeform2}) and (\ref{eq:poseq}) we 
see $ u'(t) > 0 $.  
Since $t > 0$ was arbitrary,  the proof is complete. 
\hfill$\diamondsuit$\medskip

\begin{corollary}
The solution to the initial value problem (\ref{eq:gen1})-(\ref{eq:gen2}) monotonically
approaches its steady state value $u=1$.  
\end{corollary}
\paragraph{Proof.} 
The proof follows from applying Theorem 3.1 to equation (\ref{220}).
\hfill$\diamondsuit$\medskip

\section{Related Aspects of the Newtonian Problem} 
 
To investigate the generality of the above result, consider
the nonautonomous linear damped harmonic oscillator equation
for $u=u(t)$
\begin{equation} 
u'' + bu' + u = 1 - G(t),\label{eq:sho} 
\end{equation}
as an initial value problem with arbitrary initial conditions
$u(0)$   and $u'(0)$.  We are specifically interested in the case
where $G(t) \rightarrow 0$ as $t \rightarrow \infty$, as opposed to the often
studied case where $G(t)$ is periodic (see e.g. 
\cite{hale}).  The Newtonian  sphere problem (\ref{eq:non-dim1}) is
a special case of (\ref{eq:sho}), with $b =   2-\kappa$ and $G(t) = \sqrt{\kappa/\pi t}.$
Making the change of   variables $v = u-1$, we may simplify the equation
to 
\begin{equation} 
v'' + bv' + v = - G(t)\label{eq:shov} 
\end{equation} 
so that $v=0$ solves the homogeneous equation.  Note however that if  
$G(t) \neq 0$, then $v=0$ is not a solution to (\ref{eq:shov}) for
 any $t > 0$. 
We are interested in the following question:  what  conditions  
on $G(t)$ and $b$ are necessary for the solution   $v(t)$ to remain monotone,
even within the regime of instability for the homogeneous equation.

As a first step in this direction we consider the following initial value problem  
for $t\geq0:$ 
\begin{equation} 
 v'' + b \, v' + v = -\frac{A}{\sqrt{\pi(t + t_0)}}, 
\label{eq:odealg} 
\end{equation} 
where $b,A \in \mathbb{R}$ and $t_0 \geq 0$ are constants.  
The motivation for this form is to test the 
necessity of the singularity at $t=0$ in the monotonicity result of  
Section \ref{monosection}.  To ensure complex roots, 
we assume $b \in (-2,2)$.
 
Using variation of parameters one finds a particular solution of  
(\ref{eq:odealg}) to be 
 \begin{eqnarray} 
v_p(t) &=&  \frac{A}{\beta - \alpha} \left\{ \sqrt{\beta}  
 e^{\alpha (t+t_0)} \left( \text{Erfc} \sqrt{\alpha t_0} - \text{Erfc}\sqrt{\alpha(t+t_0)} \right) \right. \nonumber \\  
&& \hspace{0.35in} 
-\left.\sqrt{\alpha}   e^{\beta (t+t_0)} \left( \text{Erfc} \sqrt{\beta t_0} -  
\text{Erfc}\sqrt{\beta(t+t_0)}\right) \right\},   
\label{up} 
\end{eqnarray} 
where $\alpha$ and $\beta$ are the roots of the characteristic polynomial 
$ m^2 + b  m + 1.$ Employing the Villat function we may 
express equation (\ref{up}) as 
\[  
v_p(t) =  \frac{A}{\beta - \alpha} \left\{ \sqrt{\beta} \text{Vi}(\alpha  
t_0) e^{\alpha t} 
 - \sqrt{\alpha} \text{Vi}(\beta t_0) e^{\beta t} \right\} + A\,M(t+t_0),  \] 
where  the function $M$ is defined by
\begin{equation}  
M(t) = \frac{1}{\alpha - \beta} \left\{\sqrt{\beta} \text{Vi}(\alpha t) -  
\sqrt{\alpha} \text{Vi}(\beta t)\right\}.  
\end{equation}
In Section \ref{monosection} we proved $M$ approaches 0 monotonically for all 
$b \in (-2,2)$. The general solution to (\ref{eq:odealg}) may be expressed
as
\begin{eqnarray} 
 v(t) & = & C_1 e^{\alpha t} + C_2 e^{\beta t} + v_p(t)  \label{prefsoln}   
\\ 
  & = &    \left\{ C_1 +  \frac{\sqrt{\beta} A \text{Vi}(\alpha t_0)}{\beta - \alpha} 
\right\} e^{\alpha t} 
      + \left\{  C_2 -   \frac{\sqrt{\alpha} A \text{Vi}(\beta t_0)}{\beta - \alpha}  
\right\} e^{\beta t} \nonumber \\
  &  & 
\qquad + \, A \, M(t+t_0).  \label{fsoln} 
\end{eqnarray} 
This equation clearly demonstrates how the long term dynamics of $v(t)$ depend on
the solution of the homogeneous problem.   In particular, it shows that the solution
will retain the 
 stability properties of the homogeneous solution unless the coefficients $C_1$ and 
$C_2$ are  chosen to zero out the first  two terms in (\ref{fsoln}).  The unique choice 
of $C_1$ and  $C_2$ for this to happen are 
\begin{equation} 
   C_1 = -\frac{A\sqrt{\beta}}{\beta - \alpha} \text{Vi}(\alpha t_0) \qquad  
\text{and } \qquad 
        C_2 = \frac{A\sqrt{\alpha}}{\beta - \alpha} \text{Vi}(\beta t_0). \label{ab} 
\end{equation} 
Moreover, it is clear from (\ref{fsoln}) that the solution in this case is  
$v(t)=A\,M(t+t_0)$, with $v(0)=A\,M(t_0)$ and  
$v'(0)=A\,M'(t_0)$.  Thus, in this case, the solution is
a translate of the monotone solution.   
The coefficients $C_1$ and $C_2$ are related to the initial conditions  
$v(0)$ and $v'(0)$ via 
\begin{equation} 
  C_1 = \frac{\beta \, v(0) - v'(0)}{\beta - \alpha}  \qquad  
\text{and } \qquad   
C_2 = \frac{v'(0) - \alpha \, v(0)}{\beta - \alpha}. \label{a0b0} 
\end{equation} 
The values of $v(0)=A\,M(t_0)$ and $v'(0)=A\,M'(t_0)$ may also be  
obtained by solving equations (\ref{ab}) and (\ref{a0b0}).
 
In summary, given $b \in (-2,2)$, $A > 0$, and $t_0 \geq 0$, for the  
equation 
\begin{equation}
v'' + b \, v' + v = -\frac{A}{\sqrt{\pi(t + t_0)}}, 
\label{gensho}
\end{equation}
there exists a unique choice of initial values $v(0)=A\,M(t_0)$,  
$v'(0)=A \, M'(t_0)$ such that the solution $v(t)$ remains monotone for all  
$t > 0$. Therefore the presence of a singularity at $t=0$ in the 
nonhomogeneous term is not necessary to obtain a monotone solution. 

In light of the above analysis it becomes clear how the solution for  the 
sedimenting sphere remains monotone in its approach to terminal velocity for 
\textit{all} relevant values of $\kappa$ (i.e., sphere densities).  
From equations (\ref{eq:non-dim1})-(\ref{eq:non-dim2}) we see
that for each value of $\kappa$, equation 
(\ref{gensho}) describes the dynamics for the dimensionless velocity
$v=u-1$, with $b =2-\kappa$, $t_0=0$, and $A=\sqrt{\kappa}$. 
Moreover, for each $\kappa \in (0,4)$  we have demonstrated that
equation (\ref{gensho}) with 
$t_0=0$, $b=2-\kappa$, and $A=\sqrt{\kappa}$, has
a unique initial value for which the solution remains monotone, namely,
\begin{equation}
  v(0)= A M(0) =  \sqrt{\kappa} \, \frac{\sqrt{\beta} - \sqrt{\alpha}}{\alpha - \beta}
 =-\frac{\sqrt{\kappa}}{\sqrt{\alpha}+\sqrt{\beta}},
\label{kd}
\end{equation}
where $\alpha$ and $\beta$ denote the roots of the polynomial $m^2 + (2-\kappa)m + 1$.
However, since $\sqrt{\alpha}$ lies in the first quadrant, $\sqrt{\alpha}
+ \sqrt{\beta} > 0$, and the computation
 \[ (\sqrt{\alpha} + \sqrt{\beta})^2 = \alpha + \beta + 2 = -b + 2 = \kappa,  \]
together with (\ref{kd}), implies 
\[  v(0)= -1.  \]
In other words,
the particular relation between the parameters $A$ and $b$
decouples $v(0)$ from all parameters, so that one obtains a monotone solution for all
values of the sphere density. 

We conclude this section with a geometric interpretation of the
monotonicity result.  In particular, we focus on  the interesting 
case of (\ref{eq:odealg}) when the parameter $b\in(-2,0)$. For 
these parameter values the solution $v=0$ of the 
homogeneous problem is unstable.   We have shown that
there is a unique set of initial conditions that defines a solution to the 
nonhomogeneous problem which approaches the unstable fixed point $v=0$
monotonically, despite the surrounding instabilities. 
Thus we return to the  fundamental puzzle posed in Section \ref{asympt}: 
How is it that the nonautonomous term in  (\ref{eq:odealg}), which 
decays to zero as $t \rightarrow \infty$, can ``stabilize'' a trajectory for 
all $ t > 0$, in the sense that this solution approaches 0 while all other
trajectories diverge due to 
the  instability of the linearized problem?
The following observation resolves the puzzle. 
First, recall that  the unique initial conditions for which the 
nonhomogeneous problem remains monotone  are  defined by
\[ v(0) = A \, M(t_0) \qquad \text{and} \qquad v'(0)= A M'(t_0).  \]  
Second, note that as the  amplitude $A$ of the nonhomogeneous term
tends to zero,  the  initial conditions  
$\left(A\,M(t_0),A\,M'(t_0)\right)$  approach $(0,0)$. This 
corresponds to the initial condition starting on the unstable equilibrium
point, which is the unique initial condition for the homogeneous
problem  whose solution does not diverge. 
In other words we have a correspondence between the trajectories of the
homogeneous equation   and the nonhomogeneous equation,  which is
continuous with respect to the  parameter $A$. The monotone
solution is then the image of the unstable fixed point under this
map. 

\section{Summary and Conclusion}

In this paper we have studied the ODE model for a sphere falling 
through a Newtonian fluid.  We have proven that the equations do not admit 
oscillations, even in the transient, in agreement with   general
experimental observations.  

From our analysis it appears that the lack of oscillations is
due to a delicate balance of terms.  It is tempting
to conclude that an oscillating motion could be produced with only a
slight modification to the equations.  However it is important that
the solution  still remain bounded, and as we have shown there is
only one trajectory which is insensitive to the linear instability
($\Re(\alpha)>0$) of the homogeneous equation. Transient oscillations
of a falling sphere have been successfully modeled by King \& Waters
using an elastic constitutive model  \cite{king:ums72}, for which a
final steady state velocity is approached.  

In principle, however, one cannot simply modify the differential
equation (\ref{eq:non-dim1}) or even (\ref{eq:maineq}) to address the
oscillations of a sedimenting sphere in a micellar fluid
\cite{belmonte:soc00,jayaraman:osf00}; one must return to the full
time-dependent partial differential equation. This was indeed how
King \& Waters obtained their result for a linear  viscoelastic
constitutive model   \cite{king:ums72}, but it is not clear that this
approach will continue to be fruitful as the complexity of the
problem increases. Self-assembling wormlike micellar solutions  are
thought to have a nonmonotonic stress/shear rate relation
\cite{spenley93,porte97}, based on the existence of an apparently
inaccessible range of shear
rates \cite{porte97,cappel97}. 
It may be that the dynamics of such a nonlinear fluid requires the
spatial information inherent in the PDEs, and that the ODE reduction
discussed here is practically limited to linear models. 
 
\section*{Acknowledgments} The authors would like to thank J. P. 
Keener, H. A. Stone, B. Ermentrout, and W. Zhang for helpful
discussions.   A.~B.~acknowledges the support of the Alfred P.~Sloan
Foundation.

\appendix 
\section{Derivation of Equations (\ref{eq:non-dim1})-(\ref{eq:non-dim2})} 
\label{app}

The  equation describing the transient motion of a falling sphere is
\begin{eqnarray}
u' + u + \sqrt{\frac{\kappa}{\pi}} \int_0^t \frac{u' (s)}{\sqrt{t-s}} ds =1 \label{eq:1},
\end{eqnarray}
where $u(t)$ is the velocity of the sphere and $\kappa$ is 
a non-dimensional parameter which depends  on the relative densities
of the sphere and the fluid. 
This integro-differential equation can be converted to a second order ODE through the 
following procedure. If
$$F(t) = \int_0^t  \frac{u' (s)}{\sqrt{t-s}} ds, $$
then Abel's theorem (see e.g., \cite[\S3.7]{keener:pam00}) implies
\begin{equation}
\int_0^t  \frac{F(\tau)}{\sqrt{t-\tau}} \, d\tau = \pi \left[u(t)-u(0)\right].
\label{abthr}
\end{equation}
Multiplying (\ref{eq:1}) by $1/\sqrt{t-\tau},$ integrating, and using (\ref{abthr})
yields the equation 
\begin{eqnarray}
  \int_0^t \frac{u'}{\sqrt{t-\tau}} \, d\tau +  \int_0^t \frac{ u}{\sqrt{t-\tau}} \, 
d\tau + \pi \sqrt{\frac{\kappa}{\pi}} \left[u(t) - u(0)\right] =   
\int_0^t \frac{1}{\sqrt{t-\tau}} \, d\tau.
 \label{eq:2}
\end{eqnarray}
From (\ref{eq:1}) one observes
\begin{eqnarray}
 \int_0^t \frac{u'(\tau)}{\sqrt{t-\tau}} \, d\tau = \sqrt{\frac{\pi}{\kappa}} 
\left(1- u - u'\right).\label{eq:n}
\end{eqnarray}
Substituting this into (\ref{eq:2}) and rewriting yields
\begin{eqnarray}
u' = \left(1- 2\sqrt{\frac{\kappa t}{\pi}}\right) +\left(\kappa -1\right) u - 
\kappa u(0)+ \sqrt{\frac{\kappa}{\pi}} \int_0^t \frac{ u}{\sqrt{t-\tau}} \, d\tau. 
\label{eq:4}
\end{eqnarray}
The desired second order differential equation is now obtained by differentiating
(\ref{eq:4}). In this regard, note that the substitution $\tau=t-x^2$ implies
\begin{equation}
I(t) = \sqrt{\frac{\kappa}{\pi}} \int_0^t \frac{u(\tau)}{\sqrt{t-\tau}} \, d\tau = 
2\sqrt{\frac{\kappa}{\pi}} \int_0^{\sqrt{t}} u(t-x^2) \, dx, \label{eq:5}
\end{equation}
thus
\begin{eqnarray}
\frac{dI}{dt}&=& \sqrt{\frac{\kappa}{\pi}}\frac{u(0)}{\sqrt{t}} + 
2\sqrt{\frac{\kappa}{\pi}}\int_0^{\sqrt{t}}  u'(t-x^2) \, dx \nonumber\\
&=& \sqrt{\frac{\kappa}{\pi}}\frac{u(0)}{\sqrt{t}} +  \sqrt{\frac{\kappa}{\pi}} 
\int_0^t \frac{u'}{\sqrt{t-\tau}} \, d\tau\nonumber \\
&=&  \sqrt{\frac{\kappa}{\pi}}\frac{u(0)}{\sqrt{t}} + (1- u - u'), \label{lst}
\end{eqnarray}
where again we have used (\ref{eq:n}).

Therefore differentiating (\ref{eq:4}) and using (\ref{lst}) yields the second order 
equation
\begin{equation}
u'' = (\kappa-2) u' - u + \left[1+\sqrt{\frac{\kappa}{\pi t}}\left( u(0)-1\right)\right].
\end{equation}
Note that from (\ref{eq:1}) the initial value of $u'$ is determined by the initial
value of $u$, i.e.,  $ u' (0) = 1 - u(0)$.

Therefore, the equation describing the transient motion of the sphere is
\begin{eqnarray} 
u'' + (2-\kappa) u' + u & = & 1 + \sqrt{\frac{\kappa}{\pi  
t}}(u(0) - 1), 
\label{eq:genn1} \\ 
u(0) & = & \xi, \quad u'(0)  =  1 - \xi. \label{eq:genn2}  
\end{eqnarray} 
If the sphere starts from rest (i.e., $u(0)=0$) then the system reduces to
\begin{eqnarray} 
u'' + (2-\kappa) u' + u = 1- \sqrt{\frac{\kappa}{\pi  
\tau}},  
\nonumber \\ 
u(0) = 0, \quad u'(0) = 1,  
\nonumber 
\end{eqnarray} 
which is precisely (\ref{eq:non-dim1})-(\ref{eq:non-dim2}).

\noindent {\sc Andrew Belmonte} \\
The W. G. Pritchard Laboratories \\
Department of Mathematics \\
Penn State University \\
University Park, PA 16802 USA \\ 
e-mail: belmonte@math.psu.edu 
\smallskip

\noindent {\sc Jon Jacobsen}\\
The W. G. Pritchard Laboratories \\
Department of Mathematics\\
Penn State University \\
University Park, PA 16802 USA\\ 
e-mail: jacobsen@math.psu.edu 
\smallskip

\noindent {\sc Anandhan Jayaraman} \\
The W. G. Pritchard Laboratories \\
Department of Mathematics \\
Penn State University \\
University Park, PA 16802 USA \\ 
e-mail: anand@math.psu.edu

\end{document}